\documentclass[11pt]{article}

\usepackage[fleqn]{amsmath}              % equation formatting
\usepackage{amssymb}                     % some math symbols, e.g., \mathbb{R}
\usepackage{booktabs}                    % pretty tables
\usepackage[margin=1in]{geometry}        % page formatting
\usepackage[pdfencoding=auto]{hyperref}  % active links within pdf
\usepackage{lineno}                      % line numbers
\usepackage{setspace}                    % line spacing
\usepackage{float}                       % Fix figures/tables in place with H
\usepackage{multirow}                    % Allow rotated text spanning multiple rows in tables
\usepackage{titling}
\usepackage{graphicx}
\usepackage{subcaption}
\usepackage[T1]{fontenc}
\usepackage{authblk}
\usepackage[backend=biber,
            natbib=true,
            dashed=false,
            maxcitenames=2,
            maxbibnames=10,
            isbn=false,
            doi=false,
            url=false,
            defernumbers=true,
            abbreviate=false,
			style=authoryear
            ]{biblatex}
\renewbibmacro{in:}{}
\renewcommand{\and}{\quad\text{and}\quad}
\newcommand{\for}{\quad\text{for}\quad}

\newcommand{\ind}[3][1]{#2 = #1,\ldots,#3}

\newcommand{\where}{\quad\text{where}\quad}
\newcommand{\with}{\quad\text{with}\quad}
\newcommand{\diag}{\text{diag}}

\newcommand{\iid}{\overset{iid}{\sim}}

\newcommand{\bbeta}{\boldsymbol\beta}
\newcommand{\bdelta}{\boldsymbol\delta}
\newcommand{\bepsilon}{\boldsymbol\epsilon}

\newcommand{\bgamma}{\boldsymbol\gamma}
\newcommand{\blambda}{\boldsymbol\lambda}
\newcommand{\bLambda}{\boldsymbol\Lambda}
\newcommand{\bmu}{\boldsymbol\mu}

\newcommand{\bOmega}{\boldsymbol\Omega}
\newcommand{\bphi}{\boldsymbol\phi}

\newcommand{\bPsi}{\boldsymbol\Psi}
\newcommand{\bSigma}{\boldsymbol\Sigma}

\newcommand{\btheta}{\boldsymbol\theta}
\newcommand{\bxi}{\boldsymbol\xi}

\newcommand{\bzero}{\boldsymbol 0}

\newcommand{\bb}{\boldsymbol b}

\newcommand{\be}{\boldsymbol e}
\newcommand{\ff}{\mbox{f}}
\newcommand{\bff}{\mbox{\bf f}}
\newcommand{\bF}{\mbox{\bf F}}
\newcommand{\bG}{\boldsymbol G}
\newcommand{\bg}{\mbox{\bf g}}
\newcommand{\bI}{\mbox{\bf I}}
\newcommand{\bR}{\mbox{\bf R}}
\newcommand{\bu}{\boldsymbol u}
\newcommand{\bv}{\boldsymbol v}

\newcommand{\bx}{\boldsymbol x}
\newcommand{\bX}{\mbox{\bf X}}

\newcommand{\bY}{\mbox{\bf Y}}

\begin{document}

%\begingroup
%\singlespacing
\title{MIMIX: a Bayesian Mixed-Effects Model for\\Microbiome Data from Designed Experiments}
\author[1]{Neal S. Grantham}
\author[1]{Brian J. Reich}
\author[2]{Elizabeth T. Borer}
\author[1]{Kevin Gross}
\affil[1]{Department of Statistics, North Carolina State University, Raleigh, NC}
\affil[2]{Department of Ecology, Evolution, and Behavior, University of Minnesota, St. Paul, MN}

\maketitle

\begin{abstract}
Recent advances in bioinformatics have made high-throughput microbiome data widely available, and new statistical tools are required to maximize the information gained from these data. For example, analysis of high-dimensional microbiome data from designed experiments remains an open area in microbiome research. Contemporary analyses work on metrics that summarize collective properties of the microbiome, but such reductions preclude inference on the fine-scale effects of environmental stimuli on individual microbial taxa. Other approaches model the proportions or counts of individual taxa as response variables in mixed models, but these methods fail to account for complex correlation patterns among microbial communities. In this paper, we propose a novel Bayesian mixed-effects model that exploits cross-taxa correlations within the microbiome, a model we call MIMIX (MIcrobiome MIXed model). MIMIX offers global tests for treatment effects, local tests and estimation of treatment effects on individual taxa, quantification of the relative contribution from heterogeneous sources to microbiome variability, and identification of latent ecological subcommunities in the microbiome. MIMIX is tailored to large microbiome experiments using a combination of Bayesian factor analysis to efficiently represent dependence between taxa and Bayesian variable selection methods to achieve sparsity.
We demonstrate the model using a simulation experiment and on a 2x2 factorial experiment of the effects of nutrient supplement and herbivore exclusion on the foliar fungal microbiome of {\it Andropogon gerardii}, a perennial bunchgrass, as part of the global Nutrient Network research initiative.

\noindent {\bf Keywords:} continuous shrinkage prior; factor analysis; microbiome; mixed model; Nutrient Network; OTU abundance data.
\end{abstract}
%\endgroup

\newpage
\section{Introduction}

A microbiome is a community of microorganisms and their genomes that belong to a particular ecological niche such as the human gut, soil, plants, or ambient dust.
Samples collected from these habitats invariably contain thousands of archaea, bacteria, and fungi which may be identified through their DNA with next-generation sequencing technologies \parencite{metzker2010sequencing}.
Understanding how these microbial communities interact with their environment holds significant implications for the fields of human health \parencite{wu2013analysis}, climate change \parencite{bond2016soil}, forensics \parencite{grantham2015fungi}, and more.
However, the tools available for characterizing microbiomes are, at present, largely limited to descriptive studies and must evolve to meet the advanced needs of the microbiome research community.
To this end, the interdisciplinary Unified Microbiome Initiative \parencite{alivisatos2015unified} aims to achieve ``predictive understanding that allows evidence-based, model-informed microbiome management and design'' by encouraging collaborative work on several promising areas of emphasis.

One such area of emphasis is the development of new statistical models for microbiome data analysis with environmental covariates, particularly in the presence of heterogeneous sources of variability.
Microbiome data are difficult to model because they are high-dimensional, sparse, over-dispersed, and possess complex dependence structure.
Moreover, as a consequence of the next-generation sequencing technology, the data are compositional, meaning they convey relative rather than absolute information; a microbe's abundance in a sample (the number of times its DNA was read by the sequencer) depends on the sequencing depth (the total number of reads).
Most standard multivariate statistical methods are designed for the analysis of absolute information and will yield spurious correlations among variables when applied indiscriminately to compositional data \parencite{pearson1896mathematical}.

In the face of these difficulties, contemporary analysis of microbiome data often works on metrics that summarize collective properties of the entire microbiome, such as measures of taxonomic diversity.
For example, in ecology, within-sample diversity is most simply measured as the mean number of unique taxa observed in a sample, referred to as $\alpha$-diversity.
Among-sample diversity, referred to as $\beta$-diversity, describes differences in taxonomic composition among samples, and may be quantified by measures like Bray-Curtis dissimilarity \parencite{bray1957ordination} or, if full taxonomic assignments are available, UniFrac distance \parencite{lozupone2005unifrac}.
Permutational multivariate analysis of variance (PERMANOVA) with pairwise differences between samples \parencite{mcardle2001fitting} is a popular tool to test whether environmental covariates are associated with significant differences in these summary metrics.
However, PERMANOVA does not yield inferences about how the environment affects individual microbes.
Additionally, implicit assumptions made by such distance-based multivariate methods may be inappropriate for ecological count data altogether \parencite{warton2012distance}.

More recently, and in a different vein, others have suggested analyzing the microbiome by fitting a separate generalized linear mixed model to the abundance of each taxon.
For instance, a linear mixed model with arcsine square root transformation or, if sparsity and overdispersion are of particular concern, a zero-inflated beta model \parencite{chen2016two} are viable methods for inferring treatment effects on the relative abundance (proportions) of taxa in the presence of random effects.  Alternative approaches model the raw abundance (counts) directly, accounting for the uncertainty associated with a taxon's abundance by conditioning on the total reads per sample \parencite{mcmurdie2014waste}. Hierarchical mixtures, such as beta-binomial and gamma-Poisson, are quite robust for this purpose and possess added flexibility for overdispersed data \parencite{zhang2017negative}.

An alternative to these univariate approaches is to model taxa within the microbiome jointly rather than individually.
%Unlike univariate models, multivariate models can potentially capture the dependence structure among microbial taxa, thus borrowing strength across taxa to detect and to estimate treatment effects.
Unlike univariate models, multivariate models can pool information across taxa to increase power for detecting and estimating treatment effects.
Towards this end, the Dirichlet-multinomial (DM) model --- the multivariate extension of beta-binomial --- provides a rich framework for modeling the entire vector of raw abundance data in each microbiome sample.
For example, the DM has proven useful for microbiome analysis in the areas of hypothesis testing and power calculations \parencite{la2012hypothesis}, sparse variable selection \parencite{chen2013variable}, and inference of microbial community structure \parencite{shafiei2015biomico}.
Despite their utility, Dirichlet variates have the undesirable property that they must negatively co-vary, making them ill-suited for modeling microbial taxa that often have positive associations, perhaps because they share similar habitat niches or because they interact symbiotically.

A more flexible dependence structure among microbial taxa is provided by \textcite{xia2013logistic}, who use a logistic normal multinomial (LNM) model that links the multinomial probability vector to a multivariate normal random variable, resulting in unconstrained occurrence probabilities on the linked scale. The covariance structure specified by \textcite{xia2013logistic} captures both positive and negative associations among taxa, unlike the DM covariance.
However, while suitable for small collections of taxa, their method for estimating this dependence structure is infeasible for high numbers of unique taxa produced by next-generation sequencing technologies.
Further, they employ a penalized regression approach to test for environmental effects, which is not conducive to uncertainty quantification.
%Dimensionality reduction would alleviate this problem, and may further help identify communities of taxa that are correlated in their collective response to environmental stimuli.
Mixed-model versions of the DM and LNM needed to analyze experiments following split-block designs have yet to be developed for microbiome data, owing to the difficulty of introducing random effects into the model hierarchy.

With these considerations in mind, we propose MIMIX (MIcrobiome MIXed model), a Bayesian mixed-effects model for analyzing microbiome data as a response variable in designed experiments.
MIMIX achieves four scientific objectives: (1) global tests of whether experimental treatments affect microbiome composition; (2) local tests for treatment effects on individual taxa and estimation of such effects if present; (3) quantification of how different sources of variability contribute to the microbiome heterogeneity; and (4) characterization of latent structure in the microbiome, which may suggest ecological subcommunities.
MIMIX is a LNM mixed model that uses Bayesian factor analysis \parencite{rowe2002multivariate} to capture complex dependence patterns among microbial taxa.
Specifically, MIMIX models high-dimensional relationships among the transformed  abundance probabilities of individual taxa through a set of low-dimensional unobservable variables, or factors.
MIMIX naturally identifies clusters of microbes that respond similarly to experimental conditions by developing continuous shrinkage Dirichlet-Laplace priors \parencite{bhattacharya2015dirichlet} for these latent factors.
We then apply Bayesian variable selection priors for the fixed effects on subpopulation abundance, reflecting the prior belief that treatments will not affect all ecological communities.
In this paper, these objectives and features of MIMIX are motivated by a multi-location randomized complete block design (RCBD) experiment that seeks to identify the influence of nutrient supplement and herbivory on the foliar fungal microbiome of a common perennial prairie bunchgrass.

The paper proceeds as follows. Section \ref{sec:data} introduces the
data that motivate our development of MIMIX in Section \ref{sec:methods}.
Section \ref{sec:sim} demonstrates
our method on simulated data in comparison with competing microbiome data
analysis methods. Finally, we apply MIMIX to RCBD experiment data in Section \ref{sec:analysis}
and close with a discussion in Section \ref{sec:discussion}. Details of posterior sampling schemes are left to the Appendix, and open-source code to reproduce the statistical analyses in this paper is available online at \url{https://www.github.com/nsgrantham/mimix}.

\section{Motivating Data}\label{sec:data}

\begin{figure}[!b]
    \centering
    \includegraphics[width=0.85\textwidth]{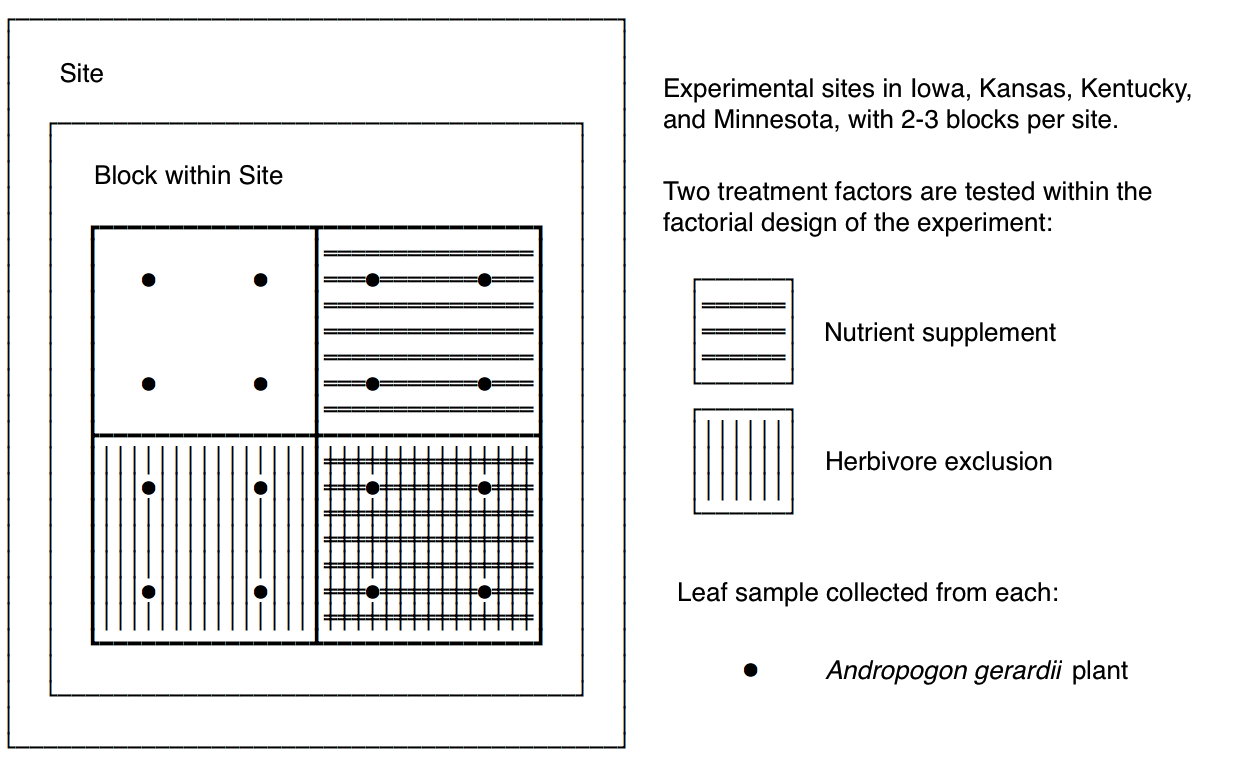}
    \caption{A schematic representation of the Nutrient Network experimental design.  This experiment replicates a 2x2 RCB design across several sites. Four plants are sampled from each plot and a microbial
    sample is collected from each plant late in the growing season (August).}
    \label{fig:experiment}
\end{figure}

The Nutrient Network (NutNet, \url{www.nutnet.org}) is a global research cooperative hosted at the University of Minnesota that uses a standardized experimental protocol to study the impact of human activity at over 100 grassland sites spanning 6 continents \parencite{borer2014finding}.  This article is motivated by data collected at 4 of these sites, all in the central US (Iowa, Kansas, Kentucky, and Minnesota). Each of these 4 sites features a 2 $\times$ 2 factorial experiment that crosses a nutrient-supplement treatment (i.e., fertilization) with an herbivore-exclusion treatment in a randomized complete block design (RCBD) (\autoref{fig:experiment}).  Here, we consider the effect of the two experimental treatments on the foliar fungal microbiome of {\it Andropogon gerardii}, a perennial bunchgrass found at each site, and native to prairie ecosystems of central North America.

NutNet scientists collected and prepared these data as follows.  In August 2014, leaf samples were collected from four {\it A. gerardii} individuals in each treatment plot. Samples were collected from plots in three replicated blocks, except in Iowa where {\it A. gerardii} was present in only two blocks. Ten samples were later removed from the study due to errors during sequencing, leaving a total of 166 samples.  Once collected, leaves were stored at 4 $^\circ$C until surface sterilization.  All leaves were surface sterilized within 24 hours after collection by exposing them to 1 minute each of 5\% sodium hypochloride (bleach), 75\% ethanol, and sterile distilled water.  Leaves were then stored at -80 $^\circ$C until DNA extraction.

The microbiome from each sample was characterized using a standard metagenomics approach.  Each surface-sterilized leaf was ground in liquid nitrogen, after which total genomic DNA was extracted using the Qiagen Plant Mini Extraction Kit (Qiagen). Genomic DNA was standardized  to 20 ng/$\mu$L. The Roche FastStart High Fidelity Taq (Roche, Indianopolis, IN, USA), described in \textcite{nguyen2016funguild}, was used to generate fungal genomic libraries.

From each individual sample, a 250 bp ITS1 region of fungal rDNA was amplified and sequenced.  Unique 7-bp sequences derived from NextGen Sequencing were used to barcode samples, as described in \textcite{smith2014sequence} and \textcite{nguyen2016funguild}. Triplicate PCR reactions were done with annealing temperatures of 51, 53 and 55 $^\circ$C, and amplicons were pooled for each sample, purified using the QIAQuick Purification Kit, and quantified using a Qubit fluorometer and Quant-iT\textsuperscript{\textregistered} dsDNA HS  Assay kit (Thermo Fisher). Equimolar concentrations of purified libraries (25ng) were pooled and sequenced in Illumina MiSeq at the University of Minnesota Genomics Center (UMGC).

Operational taxonomic units (OTUs) were identified by clustering sequences with 97\% similarity.  All samples were simultaneously analyzed using a metagenomic pipeline adapted from \textcite{nguyen2016funguild}.  Distal priming/adapter sites were removed with cutadapt v1.7.1 \parencite{martin2011cutadapt}, and reads were trimmed using Trimmomatic v0.32 \parencite{bolger2014trimmomatic}.  All short sequences or those with ambiguous bases were removed using the {\tt screen.seqs} method of Mothur v.1.34.4 \parencite{schloss2009introducing}. All remaining sequences were binned into operational taxonomic units (OTUs) with 97\% similarity using a pipeline adapted  from Song (\url{https://github.com/ZeweiSong/FAST}).

Preliminary OTU assignments were performed using the UNITE fungal database v7.0 with BLAST v2.2.28+ \parencite{camacho2009blast+}. Poor BLAST alignments (hit length $<$80\% of the alignment length, or percent identity $<$80\%) were removed. OTUs assigned to \textit{Incertae sedis} at the class level were reassigned based on taxonomic information at other levels; those that could not be reassigned were categorized as ``unidentified.''  We emphasize that these OTU assignments are preliminary and are subject to revision as reference libraries are updated, and as fungal phylogeny is more deeply understood.  Accordingly, specific ecological findings presented in this article are also preliminary, and the definitive ecological analyses of these data will be provided in forthcoming work.  Nevertheless, these data are suitable for fueling the methodological advances that we describe here.  Overall, 2,662 fungal OTUs were identified across the 166 leaf samples. Samples harbored an average of 200 unique OTUs, and many OTUs were rare, as 85\% of OTUs were identified in $<$10\% of samples.

Given the preliminary OTU assignments, we wish to investigate each of the following using these data. First, we seek to characterize how the experimental treatments affect microbiome communities. We perform this analysis in stages: first at a global level where the response is the composition of the microbiome community as a whole, and then (if the global test identifies a significant treatment effect) at a local level that characterizes the effects on the relative abundance of individual OTUs.  Second, we wish to characterize how the residual variation in the microbiome composition varies among blocks within sites and across sites, as quantifying these sources of variation may suggest insight into the ecological processes controlling these microbial communities. Finally, we wish to characterize the dependence structure among OTUs, and identify clusters of OTUs that may suggest underlying ecological subcommunities. %These objectives motivate our development of MIMIX, a robust mixed-effects model for microbiome data analysis.

\section{Methods}\label{sec:methods}

Let $Y_{ik}$ denote the count for sample $i=1,\ldots,n$ and taxon $k=1,...,K$, and let $m_i=\sum_{k=1}^K Y_{ik}$ be the total counts for sample $i$. The value of $m_i$
is an artifact of the sequencing depth of the high-throughput sequencing process
and thus analyses are performed conditional on its value.
For observation $i$, let $\bx_i$ be a $p$-vector of covariates and
let $z_i\in\{1,\ldots,q\}$ record the source of the random effects from
one of $q$ blocking factors. This latter notation may be generalized to
accommodate arbitrarily complex blocking designs, but for notational simplicity
we assume a single blocking factor in this initial development.

A multinomial likelihood is natural for multivariate count data, so we take
$\bY_i = (Y_{i1},...,Y_{iK})' \sim \mbox{Multinomial}(m_i,\bphi_i)$
where $\bphi_i = (\phi_{i1},...,\phi_{iK})'$ is the vector of expected proportions
with $\bphi_i\in\mathbb{S}^K=\{(\phi_1,\ldots,\phi_K)':\phi_k\geq 0,\ind{k}{K}, \sum_{k=1}^K\phi_k=1\}$.
We define sample-specific
$\btheta_i = (\theta_{i1}, \ldots, \theta_{iK})' \in \mathbb{R}^K$ mapped to
$\mathbb{S}^K$ by the inverse log-ratio
transformation \parencite{aitchison1986statistical}
\begin{linenomath*}
\begin{equation}\label{eq:invlogratio}
\phi_{ik} = \phi_k(\btheta_i) = \frac{\exp(\theta_{ik})}{\sum\limits_{l=1}^K\exp(\theta_{il})}\for\ind{k}{K}.
\end{equation}
\end{linenomath*}
There is a loss of dimension in transforming from $\mathbb{R}^K$ to $\mathbb{S}^K$
due to the latter's unit-sum constraint, so to achieve identifiability we restrict
each $\theta_{ik}$ to have prior mean zero.

In the spirit of \textcite{billheimer2001statistical}, we associate fixed and
random effects with the microbiome composition through the mean of $\btheta_i$.
The mixed effects decomposition is given by
$\btheta_i = \bmu + \bbeta\bx_i + \bgamma_{z_i} + \bepsilon_i$,
where $\bmu = (\mu_1,\ldots,\mu_K)'$ is the overall population mean,
$\bbeta$ is a $K\times p$ matrix of unknown fixed effect coefficients,
$\bgamma_r$ is a $K$-vector of random effects from block $r$, and
$\bepsilon_i \iid N_K(\bzero, \bOmega)$ is sample-specific random variation.

The number of taxa, $K$, is often very large in microbiome compositions.
To address complications due to high dimensionality and to account for relationships among taxa, we use Bayesian factor analysis \parencite{rowe2002multivariate}
to model the fixed and random effects within a lower dimensional representation.
For a number of factors $L$, let $\bLambda = \left(\blambda_1,\ldots,\blambda_L\right)$ be the $K\times L$
latent factor loading matrix, unknown and to be estimated. Suppose $\bLambda$ is
common to all fixed and random components of the model, a rather strong assumption, but one
that allows each latent factor $\blambda_l$, $\ind{l}{L}$, to capture sets of taxa
correlated in their response to the model's covariates and other sources of variability.
Let $\bbeta = \bLambda\bb$, $\bgamma_r = \bLambda\bg_r$, $\ind{r}{q}$, and
$\bepsilon_i = \bLambda\be_i + \bdelta_i$, $i=1,\ldots,n$.
Thus, we may represent $\btheta_i = \bmu + \bLambda\bff_i + \bdelta_i$ where
$\bff_i = \bb\bx_i + \bg_{z_i} + \be_i$ is the low-dimensional factor score
vector for sample $i$.

A prior on $\bLambda$ should ensure that the factor loading matrix captures common,
cross-species covariance that lends itself to post-hoc inference of collective taxa responses.
For instance,
setting entire columns of $\bLambda$ to zero is a means of selecting the number of active factors and setting individual
elements within $\bLambda$ to zero allows the factors to represent subsets of
taxa \citep{carvalho2008high}. To achieve both forms of sparsity, we place
continuous shrinkage priors on the high-dimensional factor loadings
$\blambda_l, \ind{l}{L}$ comprising $\bLambda$.
In particular, we select a Dirichlet-Laplace prior
\parencite{bhattacharya2015dirichlet} for its ability to detect sparse signals in
high-dimensional linear regression, which we modify here for factor analysis.
For factors $\ind{l}{L}$, let
$\blambda_l \sim \text{DL}_{a_l}$ represented by
\begin{linenomath*}
$$
\lambda_{kl}\mid\xi_{kl},\tau_l\sim\text{Lap}\left(\xi_{kl}\tau_l\right),
\;\bxi_l=(\xi_{1l},\ldots,\xi_{Kl})' \sim \text{Dir}(a_l,\ldots,a_l),\;\text{and}\;\tau_l\sim\text{Gam}(Ka_l, \nu)
$$
\end{linenomath*}
for $\ind{k}{K}$, where $\nu\sim\mathcal{G}(c_0, d_0)$ and each $a_l$ is given a discrete uniform prior over $(0,1)$ with smaller values favoring aggressive shrinkage of terms toward zero.
The Laplace distribution may be equivalently represented as a scale mixture of normals with exponential mixing density,
\begin{linenomath*}
$$
\lambda_{kl}\mid\psi_{kl},\xi_{kl},\tau_l\sim N\left(0, \psi_{kl}\xi_{kl}^2\tau_l^2\right)\and \psi_{kl}\sim\text{Exp}(1/2),
$$
\end{linenomath*}
a form that allows for straightforward Gibbs sampling of the associated parameters.

Another aim of this model is to test and quantify treatment effects on each taxon, so we place a spike-and-slab prior on $\bb$ for the purposes of stochastic variable selection \parencite{mitchell1988bayesian}.
Unlike the DL prior, the spike-and-slab prior places probability on the coefficients being exactly zero.
This allows us to compute posterior probabilities that effects are zero leading to a Bayesian test, satisfying one of MIMIX's objectives.
Let $\omega_{jl}$ be an indicator variable taking the value 1 (0) when $b_{jl}$ is included (excluded) from the model.
The spike-and-slab prior is given by $Pr(\omega_{jl} = 0) = 1-\pi_j$ and $b_{jl}\mid\omega_{jl}=1  \sim N(0, \sigma^2_b)$, where $\pi_j\sim\text{Beta}(a_0, b_0)$ is the inclusion probability for covariate $j$ in the model.
We select $a_0$ and $b_0$ such that the prior inclusion probability for each covariate is set at some $c\in[0,1]$.
In particular, the number of factors for which fixed effect $j$ is active, $S_j = \sum_{l=1}^{L}\omega_{jl}$, follows a beta-binomial distribution such that $Pr(S_j>0)=1-Pr(S_j=0)=1-\frac{\Gamma(L+b_0)\Gamma(a_0+b_0)}{\Gamma(L+a_0+b_0)\Gamma(b_0)}$, where $\Gamma(\cdot)$ is the gamma function.
Fixing this quantity at $c$ and selecting $a_0=1$ for convenience, we use the property that $\Gamma(n+1)=n\Gamma(n)$ for any positive integer $n$ to arrive at $b_0 = (\frac{1 - c}{c})L$.
A choice of $c=0.5$ reflects no prior knowledge on covariate inclusion.

Placing priors on the remaining parameters and hyperparameters of this model completes the Bayesian specification.
We use continuous priors for the intercept and random effects terms as we do not intend to test whether or not they are zero.
Let $\mu_k \iid N(0, \sigma_{\mu}^2)$, $g_{rl} \iid N(0, \sigma^2_g)$, $e_{il} \iid N(0, \sigma^2_e)$ where $\sigma^2_e = 1$ to identify the scale of $\bLambda$, and $\delta_{ik} \overset{ind}{\sim} N(0, \sigma^2_k)$ where $\sigma^2_k$ can capture over-dispersion in the sequence reads of taxon $k$.
These choices induce covariance
\begin{linenomath*}
$$
Cov(\btheta_i, \btheta_{i'}) =
\begin{cases}
  (\sigma^2_g + 1)\bLambda\bLambda' + \bSigma & i=i'\\
  \sigma^2_g\bLambda\bLambda' & i\neq i'\;\text{and}\; z_i = z_{i'} \\
  \bzero & z_i \neq z_{i'}
\end{cases}
$$
\end{linenomath*}
where $\bSigma$ is the $K\times K$ diagonal matrix with diagonal elements
$(\sigma^2_1, \ldots, \sigma^2_K)$.
We suppose the variance terms of the model follow independent inverse gamma priors with shape $u_0$ and scale $v_0$.

Posterior sampling is conducted using Markov chain Monte Carlo (MCMC). Most terms in this model
formulation are conjugate and are updated via Gibbs sampling. One exception is
$\btheta_i$ which we update with Hamiltonian Monte Carlo (HMC)
\parencite{neal2011mcmc}. The details of these sampling schemes are found in the
Appendix. Code to perform MCMC is written in Julia \parencite{bezanson2017julia} and available online at \url{https://www.github.com/nsgrantham/mimix}.

\section{Simulation Study}\label{sec:sim}

We test our model on a simulated experiment with $K=100$ taxa, $p=1$ fixed effect, one blocking factor that takes $q=5$ levels, and $n=40$ observations with $8$ assigned to each block.
Within each block exists a balanced experiment with two levels of a single experimental factor, where $x_i=1$ if observation $i$ receives one level of the experimental factor, and $x_i=0$ otherwise, and $z_i=r$ if observation $i$ belongs to block $r$.

The fixed treatment effect, $\bbeta$, is a sparse $K$-vector with a varying percentage
of non-zero elements: 0\% dense ($\bbeta = \bzero$, i.e., no signal), 5\% dense, or 10\% dense.
To generate $\bbeta$ that is 5\% dense, we partition the taxa
into 20 clusters of 5 taxa each, select one of the twenty groups at
random, and draw a value from $v\sim\text{Unif}\left([-3, -1)\cup(1, 3]\right)$
to signify the group's collective response to the fixed treatment. That is,
$\beta_k=v$ if taxon $k$ belongs to the selected group and $\beta_k=0$ otherwise.
For 10\% dense, we repeat this process once more on a second group.
These groups are designed to represent taxa with shared phylogenetic ancestry or taxa that react similarly to the fixed effect.

For the random blocking effects, we draw $\bgamma_r \iid N_K(\bzero, \bSigma_\gamma)$ with
autoregressive covariance $(\bSigma_\gamma)_{kk'} = \sigma^2_{\gamma}\rho_\gamma^{|k-k'|}$, $\rho_{\gamma}=0.9$.
Block-to-block variability, $\sigma^2_{\gamma}$, is set at 1 (medium) or 4 (high).
For each observation $i$, define $\btheta_i = \bmu + \bbeta x_i + \bgamma_{z_i} + \bepsilon_i$,
where $\bmu$ is a vector of length $K$ with equally-spaced steps from 1 to -1 and $\bepsilon_i\iid N_K(\bzero,\bSigma_{\epsilon})$ with autoregressive covariance
$(\bSigma_{\epsilon})_{kk'} = \sigma^2_{\epsilon}\rho_{\epsilon}^{|k-k'|}$. We fix
$\rho_{\epsilon}=0.9$ and examine sample-to-sample variability,
$\sigma^2_{\epsilon}$, over 1 (medium), 4 (high), and 9 (very high).
Finally, we arrive at the final data by drawing each vector of counts $\bY_i$
from a multinomial distribution with total counts $m_i$ chosen at-random from
$2,500$ to $5,000$ and taxa proportions $\bphi_i$ calculated according to \eqref{eq:invlogratio}.
We do not generate data directly from the MIMIX model to test for robustness to model assumptions (e.g., fixed and random effect correlation); data generated assuming a low-dimensional dependence structure would unduly favor MIMIX over the competitors.

In total, we examine each of 18 factor combinations (0, 5, and 10 \% dense,
medium/high block variance, medium/high/very high error variance) with 50 replications and
compare the performance of three competing microbiome data analysis methods:

\begin{enumerate}
  \item \textbf{PERMANOVA}: Permutational multivariate analysis of variance \parencite{mcardle2001fitting},
    or PERMANOVA, with Bray-Curtis dissimilarity (BC), a common analysis procedure in ecology. This method tests the null hypothesis that there is no difference
    between the centroids of the treatment and control groups in BC space, but is
    unequipped to perform parameter estimation. It is implemented by {\tt adonis}
    in the {\tt R} package {\tt vegan 2.3-5} {\tt R}, among other available software options.
\item \textbf{MIMIX}: Our Bayesian mixed-effects model as presented in Section~\ref{sec:methods} with $L=40$ factors, the number of observations.
      10,000 posterior samples are collected with 5,000 removed for burn-in.
  \item \textbf{MIMIX w/o Factors}: Bayesian mixed-effects model with no factors. This formulation mimics
    the available mixed model approaches to microbiome data analysis that do not account for dependence patterns among taxa, i.e.,
    $\bLambda = \bI_K$ and $\be_i = \bzero$ for all $i=1,\ldots,n$.
    10,000 posterior samples are collected with 5,000 removed for burn-in.
 \end{enumerate}

\begin{figure}[!b]
    \centering
    \includegraphics[width=1\textwidth]{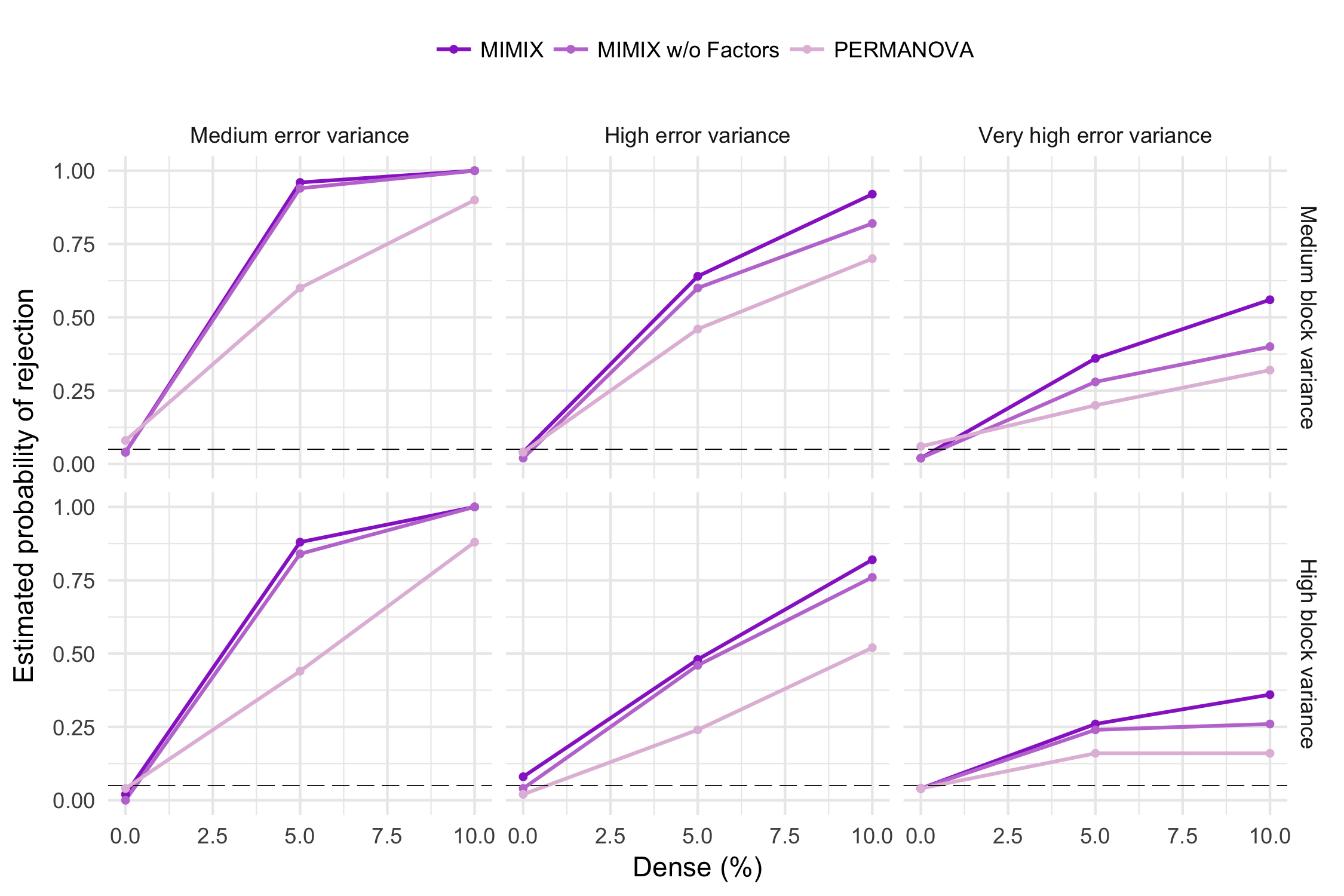}
    \caption{Results of a global test for treatment effect by MIMIX, MIMIX w/o Factors, and PERMANOVA,
        under a variety of simulation conditions. When $\bbeta = \bzero$ (0\% Dense) the line gives the test's type I error, where a dashed line at 0.05 is provided for reference.
        For $\bbeta \neq \bzero$ (>0\% Dense), the line values depict the statistical power of each test.
        }\label{fig:power}
\end{figure}

The methods are first evaluated on their power/type I error of a global test for treatment effect where PERMANOVA rejects for $p$-value $<0.05$ and MIMIX and MIMIX w/o Factors reject if $Pr(\bbeta\neq\bzero\mid \bY) > 0.9$. Both MIMIX and MIMIX w/o Factors regularly outperform PERMANOVA in detecting the presence of a significant signal (Figure \ref{fig:power}). In situations with medium error variance, MIMIX and MIMIX w/o Factors achieve similar power regardless of the block variance. As error variance increases, MIMIX is more likely than MIMIX w/o Factors to correctly identify the presence of a significant treatment effect. In most scenarios, the type I error of these methods falls at or below 0.05.

We further compare MIMIX and MIMIX w/o Factors on their local tests and estimation of
treatment effects for each OTU. Several metrics are considered:
root mean squared error, $\text{RMSE} = \sqrt{\frac{1}{K}\sum_{k=1}^K(\hat{\beta}_k^{\text{mean}}-\beta_k)^2}$
where $\hat{\beta}_k^{\text{mean}}$ is the posterior mean of $\beta_k$,
coverage of 95\% credible intervals, $\text{C95} = \frac{1}{K}\sum_{k=1}^KI(\hat{\beta}_k^{0.025} < \beta_k < \hat{\beta}_k^{0.975})$
where $\hat{\beta}_k^{q}$ is the posterior $q$th quantile of $\beta_k$,
and the true positive rate (TPR) and true negative rate (TNR)
of local tests that reject if the 95\% credible interval for $\beta_k$ excludes zero. Table \ref{tab:simulation} gives the values of these metrics averaged over 50 replications at each factor combination.

% latex table generated in R 3.3.1 by xtable 1.8-2 package
% Thu Feb  2 13:02:01 2017
\begin{table}[!t]
\caption{Local test and estimation performance for MIMIX and MIMIX w/o Factors under a variety of
        simulation conditions as measured by root mean squared error (RMSE), coverage of 95\% credible intervals (C95), true positive rate (TPR), and true negative rate (TNR). All values are multiplied by 100.}\label{tab:simulation}
\centering
\scalebox{0.9}{
\begin{tabular}{lllrrrrrrrr}
  \toprule
  & & & \multicolumn{4}{c}{Medium block variance} & \multicolumn{4}{c}{High block variance} \\
  \cmidrule(r){4-7} \cmidrule(r){8-11}
  Dense & Error var. & Method & RMSE & C95 & TPR & TNR & RMSE & C95 & TPR & TNR \\
  \midrule
0\%  & Medium    & MIMIX       & 2.8  & 100.0 &    - & 100.0 & 2.3 & 99.9  &    - & 99.9 \\ \vspace{0.5em}
     &           & w/o Factors & 0.9  & 100.0 &    - & 100.0 & 1.1 & 100.0 &    - & 100.0 \\
     & High      & MIMIX       & 6.0  & 100.0 &    - & 100.0 & 7.5 & 99.9  &    - & 99.9 \\\vspace{0.5em}
     &           & w/o Factors & 2.0  & 100.0 &    - & 100.0 & 3.0 & 100.0 &    - & 100.0 \\
     & Very High & MIMIX       & 7.8  & 100.0 &    - & 100.0 & 6.7 & 99.9  &    - & 99.9 \\ \vspace{1em}
     &           & w/o Factors & 2.6  & 100.0 &    - & 100.0 & 4.0 & 100.0 &    - & 100.0 \\
5\%  & Medium    & MIMIX       & 8.2  & 99.7  & 92.0 & 100.0 & 11.3 & 99.4 & 81.6 & 100.0 \\ \vspace{0.5em}
     &           & w/o Factors & 8.9  & 99.5  & 76.8 & 100.0 & 12.7 & 99.3 & 59.6 & 100.0 \\
     & High      & MIMIX       & 18.2 & 99.2  & 59.6 & 100.0 & 25.5 & 98.0 & 42.8 & 99.8 \\ \vspace{0.5em}
     &           & w/o Factors & 20.8 & 98.6  & 31.2 & 100.0 & 29.5 & 97.5 & 15.2 & 100.0 \\
     & Very High & MIMIX       & 31.2 & 97.3  & 21.6 & 100.0 & 32.8 & 97.6 & 15.2 & 99.9 \\ \vspace{1em}
     &           & w/o Factors & 34.8 & 96.5  & 6.0  & 100.0 & 36.9 & 96.8 & 4.0 & 100.0 \\
10\% & Medium    & MIMIX       & 13.0 & 99.0  & 97.0 & 99.6  & 16.9 & 98.5 & 87.8 & 99.6 \\ \vspace{0.5em}
     &           & w/o Factors & 11.9 & 99.4  & 87.4 & 99.9  & 17.2 & 98.8 & 67.2 & 100.0 \\
     & High      & MIMIX       & 28.4 & 97.8  & 49.2 & 99.8  & 35.5 & 96.1 & 39.6 & 99.6 \\ \vspace{0.5em}
     &           & w/o Factors & 35.9 & 97.1  & 24.4 & 100.0 & 42.7 & 95.8 & 12.6 & 100.0 \\
     & Very High & MIMIX       & 42.6 & 95.8  & 20.8 & 100.0 & 47.8 & 93.7 & 10.4 & 99.8 \\
     &           & w/o Factors & 51.0 & 93.9  & 7.8  & 100.0 & 52.1 & 92.7 & 1.8 & 100.0 \\
   \bottomrule
\end{tabular}
}
\end{table}

When there is no signal present in the fixed effects (i.e., $\bbeta = \bzero$, or 0\% dense), MIMIX w/o Factors achieves lower RMSE on average than MIMIX in estimating all treatment effects to be zero, and both methods yield credible intervals that nearly always correctly include zero.
In practice, these estimates are inconsequential if the global test appropriately fails to identify a significant treatment effect. When a small proportion of OTUs are affected by the treatment (5\% dense and 10\% dense), MIMIX regularly outperforms MIMIX w/o Factors in the detection and estimation of these non-zero fixed effects. Specifically, when error variance is medium, the TPR of MIMIX is very high (81.6\% to 97.0\%) and beats MIMIX w/o Factors (59.6\% to 87.4\%), whereas the RMSE of the two methods are comparable. For high error variance, the TPR drops to about 50\% (MIMIX) and 20\% (MIMIX w/o Factors), with higher RMSE achieved by both methods, as expected, but lower RMSE obtained by MIMIX on average. This trend continues with very high error variance, resulting in lower TPR at around 20\% and 5\% respectively, and greater difference in RMSE in favor of MIMIX. Both methods are strongly conservative, achieving TNRs that are overwhelmingly 100\%, and their C95 is often greater than the expected 95\%, except in extreme variance situations (very high error variance, high block variance) with 10\% dense fixed effects.

From the global and local simulation results, we draw three broad conclusions about the performance of MIMIX, MIMIX w/o Factors, and PERMANOVA for microbiome data analysis in designed experiments. First, at a global level, MIMIX and MIMIX w/o Factors achieve far greater power and comparable type I error (about 0.05) to PERMANOVA. Moreover, the global tests for all three methods appear more adversely affected by higher overdispersion in taxa counts than higher variability introduced by blocking factors in the experimental design. Second, at the local level, MIMIX is better suited for both the detection and estimation of sparse treatment effects compared to MIMIX w/o Factors.
In this case, MIMIX w/o Factors achieves lower TPR and higher RMSE on average because it does not account for correlation patterns among taxa.
Finally, TNR and C95 are very high and relatively consistent between the two methods under all simulation conditions, suggesting MIMIX and MIMIX w/o Factors are conservative in detecting significant OTU-specific fixed effects.

\section{Analysis of the NutNet Experiment}\label{sec:analysis}

We first compare the performance of MIMIX and MIMIX w/o Factors on the NutNet data through five-fold cross-validation, setting the maximum number of latent factors ($L$) for MIMIX equal to the number of samples (166).
Specifically, for each $\bY_i$ with total counts $m_i$ we construct $\bY_i^{\text{test} f}$, $f=1,\ldots,5$ by assigning each of $m_i$ observations to one of the five folds at random and let $\bY_i^{\text{train} f} = \bigcup_{g \neq f} \bY_i^{\text{test} g}$. % $\bY_i - \bY_i^{\text{test} f}$.
For each fold $f=1,\ldots,5$, we fit both models to training data $f$, drawing
20,000 posterior samples and discarding the first 10,000 for burn-in. Next, we examine
the difference of their log-likelihoods (MIMIX minus MIMIX w/o Factors) evaluated on
testing data $f$ where the
multinomial probability vector is estimated by the normalized posterior mean
vector of occurrence probabilities $\hat{\bphi}_i$. Over all five folds,
69\% of differences are positive on average, favoring MIMIX. Thus, MIMIX appears to be a more apt model for these data than MIMIX w/o Factors.

\begin{figure}[!b]
   \centering
    \includegraphics[width=1\textwidth]{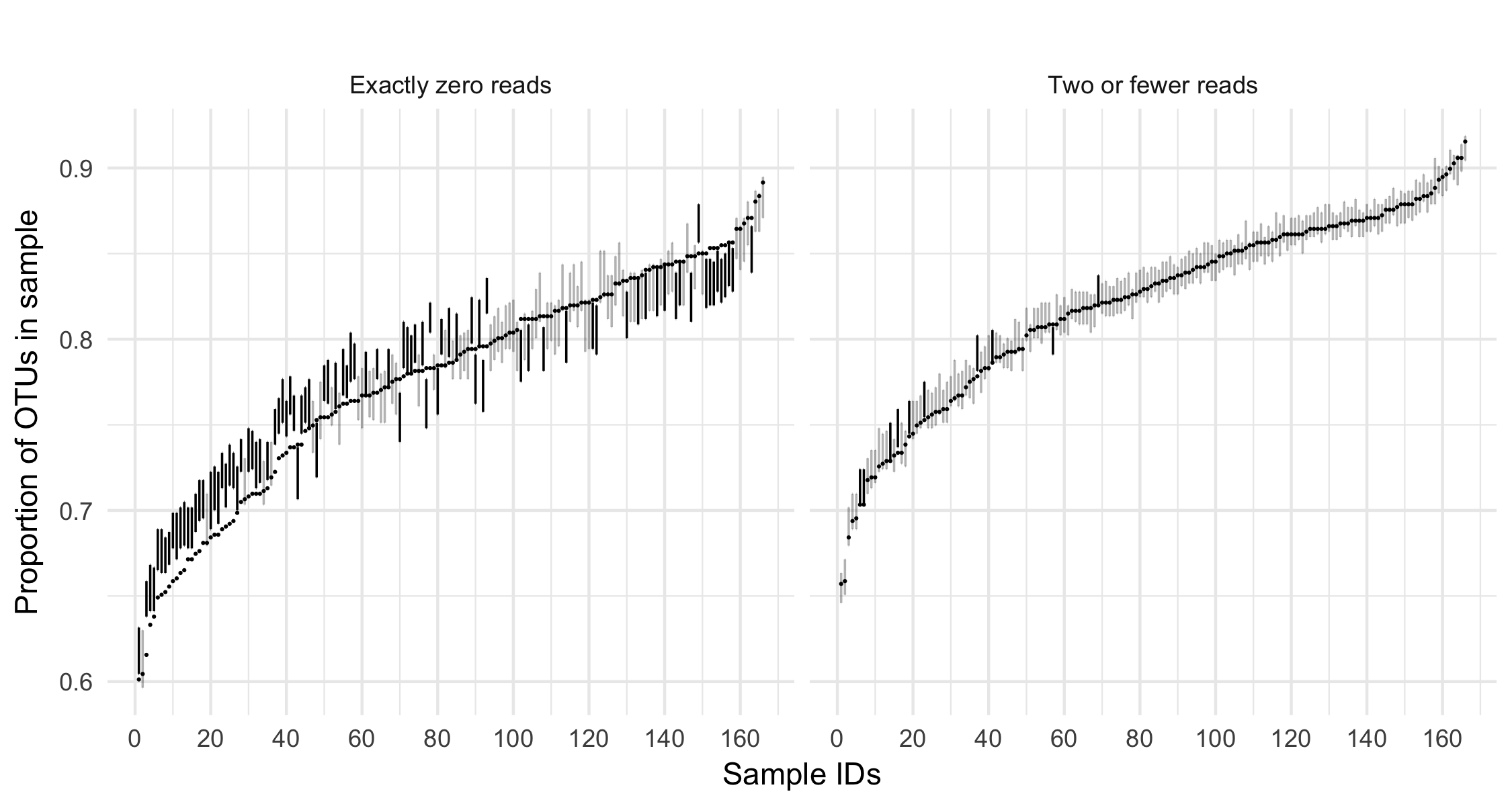}
    \caption{Posterior predictive checks on sparsity for MIMIX applied to the NutNet data. For each sample, the dot indicates the proportion of OTUs in the sample with 0 (left) or $\leq 2$ (right) reads.  The vertical line is the 95\% posterior predictive distribution, shaded black if the interval excludes the observed value and gray otherwise.}
    \label{fig:post-pred-sparsity}
\end{figure}

To assess whether MIMIX adequately captures the sparsity and overdispersion in the data, we fit a preliminary model to the data and perform posterior predictive checks \parencite{gelman2014bayesian}. These checks examine the proportion of OTUs within each sample with zero counts (sparsity) and the maximum proportion of total counts within a sample from a single OTU (overdispersion). This is done by predicting new $\bY_1,\ldots,\bY_n$ from every posterior sample and comparing the sparsity and overdispersion in these predicted samples with the observed data. With respect to sparsity, we also consider the proportion of OTUs within each sample with two or fewer counts, as these singletons and doubletons are thought by biologists to be generated by errors in the sequencing process. Figure ~\ref{fig:post-pred-sparsity} depicts posterior predictive checks on sparsity of MIMIX after 20,000 posterior samples with the first 10,000 removed for burn-in. MIMIX overestimates the proportion of OTUs within each sample with exactly zero counts, but when singleton and doubleton counts are further considered the model recovers the observed near-sparsity of the original data. The distinction between zero counts and two or fewer counts is likely of little consequence. Posterior predictive checks on overdispersion with MIMIX suggest that the maximum proportions are also well captured by the model.

We now use MIMIX to characterize the effects of the nutrient-supplement and herbivore-exclusion treatments on the fungal foliar microbiome of {\it A.\ gerardii}.
For the purposes of comparison, we also present analyses from Bray-Curtis PERMANOVA, which represents the current state-of-the-art in ecological analysis, and MIMIX w/o Factors.  We present MIMIX w/o Factors to illustrate the consequences of using factor analysis to account for dependence patterns among taxa in the microbiome, but we emphasize that our simulation studies and preliminary analysis point towards MIMIX as the most trustworthy analysis. For the Bayesian models, we collect 20,000 posterior samples and discard the first 10,000 for burn-in.

First, we conduct a global test of whether the experimental treatments affect the overall composition of the microbiome. No method identifies a significant interaction effect between the two treatments, with PERMANOVA $p=0.120$, MIMIX posterior probability $0.667$, and MIMIX w/o Factors posterior probability $0.413$.
However, PERMANOVA and MIMIX find strong evidence that the fungal microbiome composition is affected by nutrient supplement, with $p= 0.003$ and posterior probability $1.0$, respectively, while MIMIX w/o Factors does not, with posterior probability $0.757$.
No method detects an effect of herbivore exclusion, with PERMANOVA $p=0.787$, MIMIX posterior probability $0.505$, and MIMIX w/o Factors posterior probability $0.523$.
These results suggest that the composition of the foliar fungal microbiome of {\it A. gerardii} is impacted by the resources available to the plant host.

\begin{figure}[H]
\centering
\begin{subfigure}{.48\textwidth}
  \centering
  \includegraphics[width=1\linewidth]{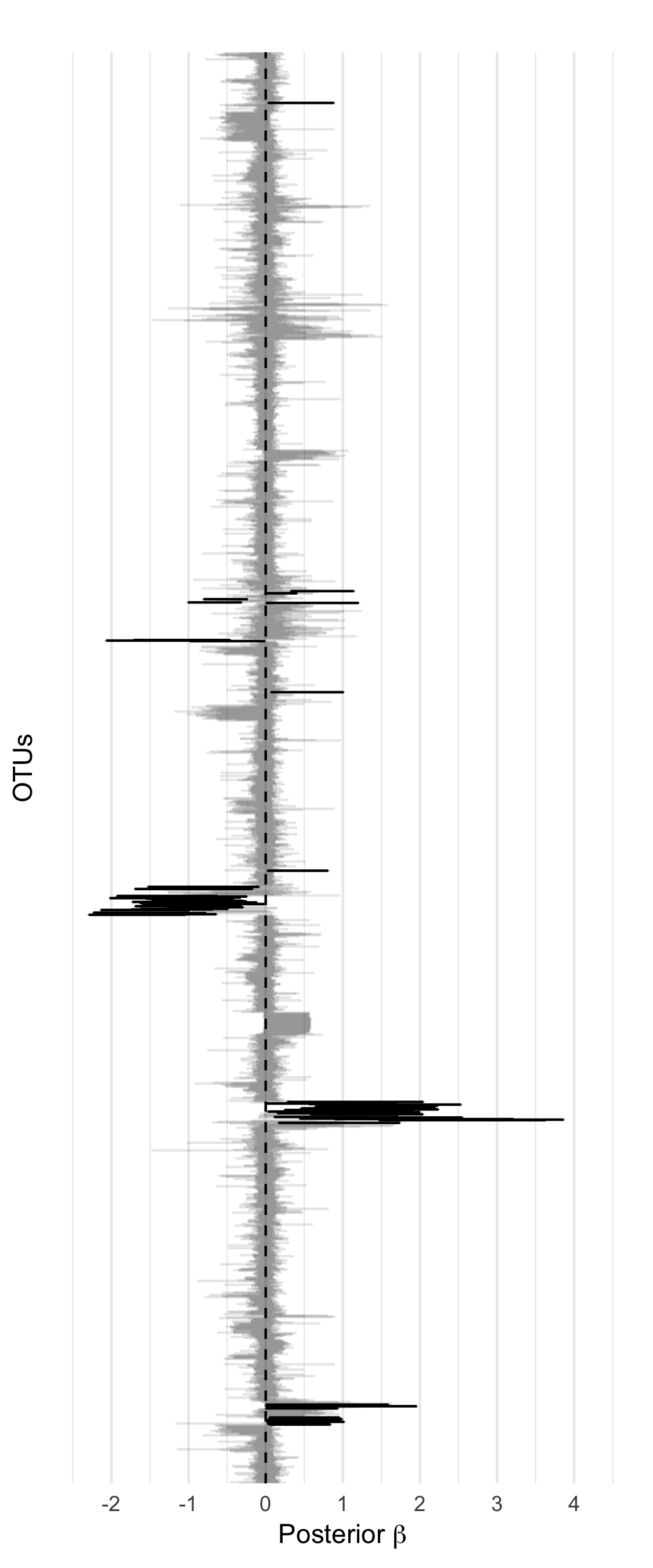}
  \caption{All OTUs}
  \label{fig:beta_all}
\end{subfigure}%
\begin{subfigure}{.48\textwidth}
  \centering
  \includegraphics[width=1\linewidth]{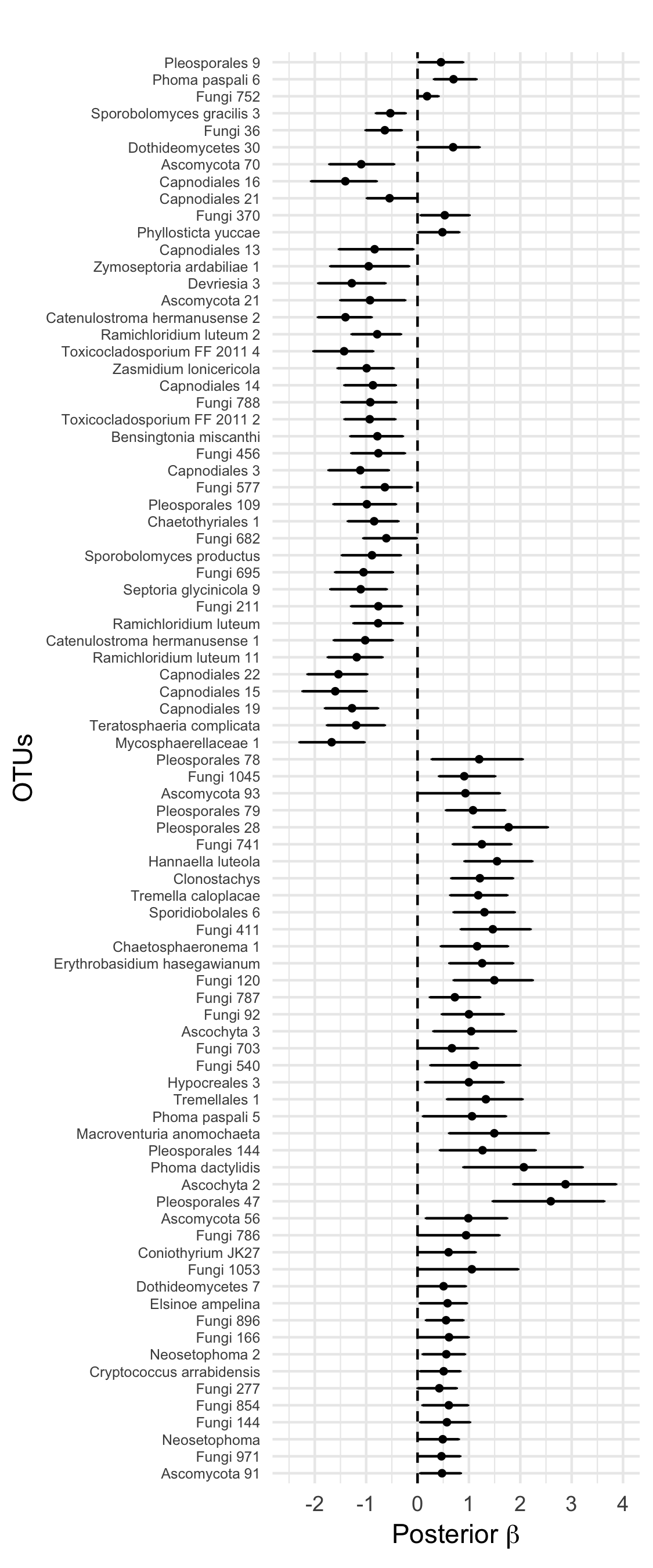}
  \caption{Only affected OTUs}
  \label{fig:beta_signif}
\end{subfigure}
\caption{Posterior 95\% credible intervals for the effect of nutrient supplement on each OTU. Across all OTUs (a), most are not significantly affected by the treatment (gray lines), but MIMIX identifies 84 of 2,662 OTUs (3.2\%) that show a significant response (black lines). Among these affected OTUs (b), the taxonomy of each fungal OTU is given up to species, if known, or at a higher taxonomic rank, such as genus or order, with trailing numbers identifying distinct strains.
The OTUs are ordered along the y-axis according to complete linkage hierarchical clustering of the estimated factor correlation matrix from MIMIX.}
\label{fig:beta}
\end{figure}

We next use MIMIX to estimate the effects of nutrient supplement on individual fungal OTUs.
Because the OTU assignments for this particular data set are only preliminary, we focus here on the distribution of OTU-level effects, and reserve the characterization of effects on specific OTUs for later work. The fixed effect for nutrient supplement estimated by MIMIX has 95\% credible intervals that exclude zero for 84 OTUs (Figure \ref{fig:beta}). Thus, while this analysis finds overwhelming evidence that environmental nutrient supply alters the composition of these microbiomes, this effect appears to be driven by only a few constituent microbes. Moreover, it appears accounting for correlation among OTUs is essential to detecting these individual microbes.

% \begin{figure}[H]
%     \centering
%     \includegraphics[width=1\textwidth]{betarange.png}
%     \caption{Posterior 95\% credible intervals of $\bbeta$, the effect of nutrient supplement on
%         each OTU. Most OTUs are not significantly affected by the treatment (gray lines),
%         but MIMIX identifies 84 of 2,662 OTUs (3.2\%) that show a significant response (black lines), whereas MIMIX w/o Factors finds only a single affected OTU. The OTUs
%         are ordered along the y-axis according to complete linkage hierarchical clustering of
%         the estimated factor correlation matrix from MIMIX.}
%     \label{fig:beta_all}
% \end{figure}

% \begin{figure}[H]
%     \centering
%     \includegraphics[width=1\textwidth]{betarange-signif.png}
%     \caption{Posterior means and 95\% credible intervals of $\bbeta$ from MIMIX for only those OTUs which show a
%         significant response to nutrient supplement. The taxonomy of each fungal OTU is given up to species, if known, or at a higher taxonomic rank, such as genus or order, with trailing numbers identifying distinct strains.}
%     \label{fig:beta_signif}
% \end{figure}

\begin{figure}[!t]
    \centering
    \includegraphics[width=1\textwidth]{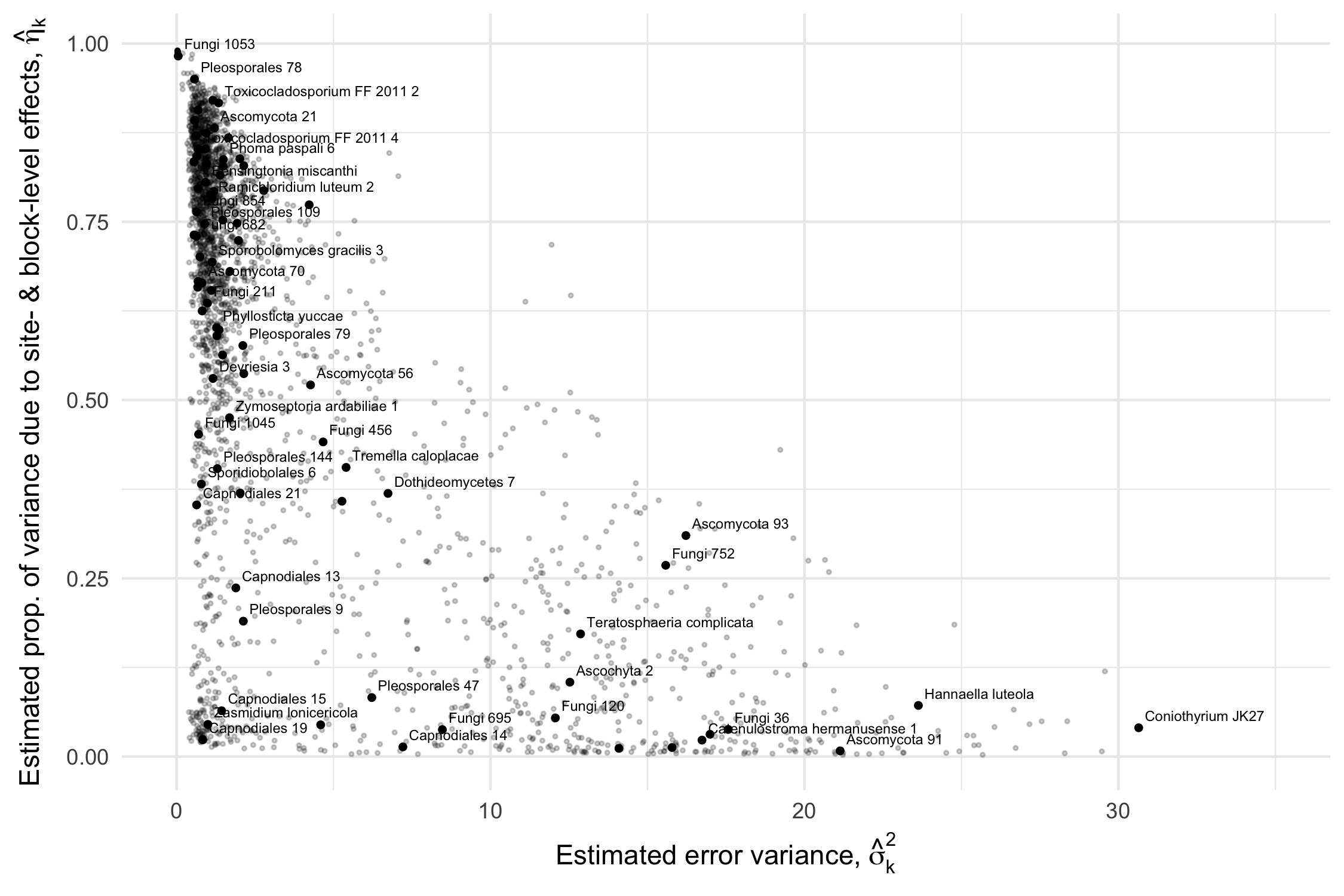}
    \caption{The proportion of variance for each OTU that is explained by the contribution of site and block vs.\ unexplained residual variation. Black dots correspond to OTUs identified by MIMIX as being significantly affected by nutrient supplement in Figure~\ref{fig:beta_signif}, though only a subset of names are displayed to avoid overlapping labels.}
    \label{fig:proportion-variance}
\end{figure}

\begin{figure}[!b]
    \centering
    \includegraphics[width=0.95\textwidth]{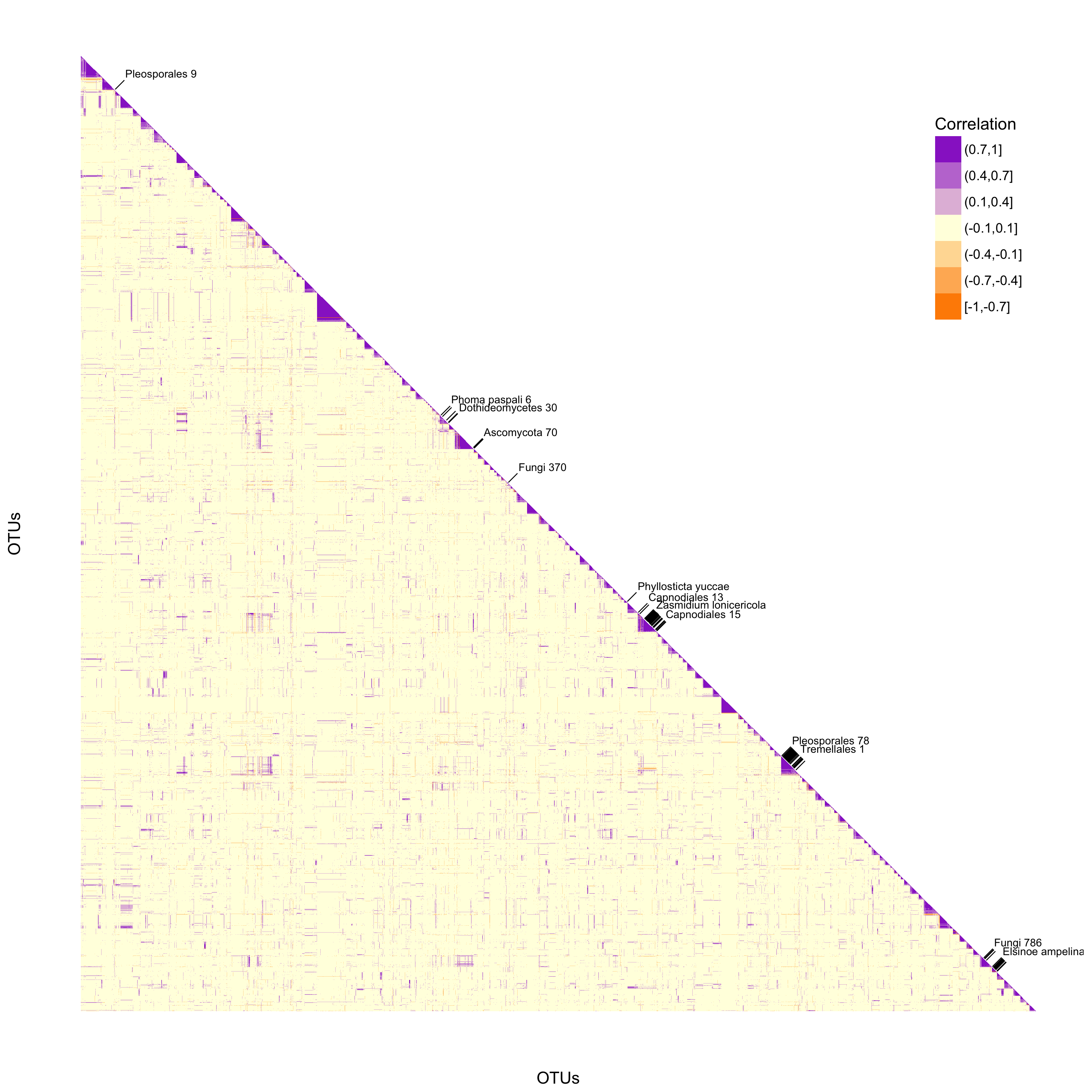}
    \caption{Estimated factor correlation matrix among OTUs, with OTUs ordered by hierarchical clustering. Clusters of strongly correlated OTUs, represented by purple triangles along the diagonal, indicate small collections of taxa that respond similarly to the fixed and random effects of the designed experiment. The OTUs highlighted along the  diagonal are those identified by MIMIX as being significantly affected by nutrient supplement in Figure~\ref{fig:beta_signif}, though only a subset of names are displayed to avoid overlapping labels.
    }
    \label{fig:correlation}
\end{figure}

MIMIX quantifies the relative contribution of different sources of residual variation to fungal composition.
Site- and block-level variances are estimated with posterior means $\hat{\sigma}^2_{\text{Site}} = 2.296$ and $\hat{\sigma}^2_{\text{Block}} = 0.355$.  Posterior means of OTU-specific variances not attributed to the study design ($\hat{\sigma}_1^2, \ldots, \hat{\sigma}_K^2$) are strongly skewed, ranging from 0.032 to 30.65 with mean 3.753 (Figure \ref{fig:proportion-variance}). The proportion of residual variance in OTU $k$ that is captured by site- and block-level effects is estimated by
\begin{linenomath*}
$$
\hat{\eta}_k =1- \frac{\hat{\sigma}_k^2}{\hat{\sigma}_k^2 + (1 +\hat{\sigma}^2_{\text{Site}} + \hat{\sigma}^2_{\text{Block}})\sum_{l=1}^L\hat{\lambda}_{lk}^2},
$$
\end{linenomath*}
where $\hat{\lambda}_{lk}$ is the posterior mean of $\lambda_{lk}$. Figure~\ref{fig:proportion-variance} shows $\hat{\eta}_k$ vs.\ $\hat{\sigma}^2_k$ for each OTU. Two loose groups of points emerge, one in which residual variation is almost entirely explained by site and block variation ($\hat{\eta}_k$ > 0.5), and another in which these random effects explain a relatively small amount of residual variation in OTUs ($\hat{\eta}_k$ < 0.2). OTUs identified by MIMIX as being significantly affected by nutrient supplement, indicated by black dots in Figure~\ref{fig:proportion-variance}, appear to be strongly represented in the former of these two groups, although there are still many OTUs that are significantly affected by nutrient supplement that do not appear to be influenced by site and block effects. Overall, site and block effects explain over half the residual variation ($\hat{\eta}_k$ > 0.5) for approximately 66\% of OTUs.

Finally, we take a closer look at the estimated factor correlation matrix $\bLambda\bLambda'$ using the posterior mean of $\bLambda$ (Figure \ref{fig:correlation}). MIMIX identifies a large number of potential clusters of fungal OTUs (grouped along the diagonal of Figure \ref{fig:correlation}), with myriad positive and negative correlations among them.  These clusters may indicate latent ecological subcommunities, or they may reflect collections of taxa that occupy similar ecological niches.  MIMIX w/o Factors does not adequately account for these relationships among OTUs, which explains its poorer fit to the data.

\section{Discussion}\label{sec:discussion}

In this paper, we introduce MIMIX (MIcrobiome MIXed effects), a Bayesian mixed-effects model to
analyze microbiome data as a response variable in designed experiments. MIMIX has several
attractive features for the analyis of high-dimensional, sparse, microbiome count data.
It performs spike-and-slab variable selection to identify treatment effects on individual
Operational Taxonomic Units (OTUs). Moreover, its Bayesian factor analysis formulation with
a continuous shrinkage Dirichlet-Laplace prior clusters OTUs into different factors based
on how they respond to the fixed and random effects in the experiment. This allows for post-hoc
analysis of the model to identify behaviorally similar clusters of OTUs within the larger microbiome community. In a simulation study, these features allow MIMIX to outperform both PERMANOVA with Bray-Curtis dissimilarity and MIMIX that does not include Bayesian factors (MIMIX w/o Factors) in identifying and estimating sparse treatment effects.

We demonstrate MIMIX on experimental data from four sites within the Nutrient Network cooperative to quantify the effects of nutrient supplement and herbivore exclusion on the fungal microbiome of the grass species \textit{Andropogon gerardii}. We identify a significant effect of nutrient supplement (but not herbivore exclusion) on these microbiomes, while accounting for random effects due to both site and blocks within site.  We also identify a significant treatment effect of nutrient supplement on about 3.2\% of OTUs. Although the OTU assignments in this particular data set are preliminary, our results illustrate how MIMIX enables OTU-level inferences that may allow for deeper and sharper understanding of how environmental conditions impact the abundance of specific taxa in a microbiome.

Ecologically, this analysis of the Nutrient Network data suggests the following insights.  First, ecologists are frequently interested in how resource supply and grazing combine to influence the structure of ecological communities (the so-called ``bottom-up'' vs.\ ``top-down'' dichotomy).  The results of this experiment suggest that resource supply, or ``bottom-up'' factors, play a larger role in structuring a host's microbiome than predation.  Second, the paucity of large OTU-specific responses (Figure \ref{fig:beta}) suggests that only a handful of microbial taxa respond to the nutrient supplementation, and that these responses can be sufficient to reshape the microbiome when considered as an ecological whole. Third, the residual variation of some, though not all OTUs can be explained by site- and block-level random effects (Figure \ref{fig:proportion-variance}), suggesting that these OTUs may either be strongly influenced by regional environmental correlates, or may be limited by reduced dispersal at regional (km) scales. Finally, the estimated factor correlation matrix (Figure \ref{fig:correlation}) suggests that this foliar microbiome is composed of many modestly sized clusters of similarly behaving OTUs.  This pattern may either suggest many moderately sized subcommunities, aggregation of taxa into many separate ecological niches, or both.

The initial results from MIMIX are encouraging, but its features will need to scale as microbiome experiments grow in complexity. For example, MIMIX is not currently suited for handling data from longitudinal studies with repeated measures over time.  Furthermore, while the dimensionality of the microbiome data analyzed here is quite high at $K\approx2,500$, the dimensionality can grow rapidly, especially when multiple domains of life (bacteria,  archaea, fungi, etc.) are studied. In such instances, computation time and memory management will become a more pressing concern which may require a reconstruction of the posterior sampling scheme.

\section*{Funding}

This work was supported by the National Science Foundation EF-1241794.

\printbibliography

\section*{Appendix}\label{sec:appendix}

\subsection*{Hamiltonian Monte Carlo (HMC) for $\btheta_i, \ind{i}{n}$}\label{ssec:HMC}

The negative log posterior distribution of $\btheta_i$ is given by
\begin{equation}\label{eq:nlp_theta}
  -\log p(\btheta_i\mid \cdots) \propto -\bY_i'\btheta_i + m_i\log\left[\sum_{k=1}^K\exp(\theta_{ik})\right] + \frac{1}{2}\sum_{k=1}^K\sigma^{-2}_k(\theta_{ik} - \mu_k - \blambda_k'\bff_i)^2
\end{equation}
with gradient
\begin{equation}\label{eq:grad_nlp_theta}
  \nabla_{\theta_i}\left[-\log p(\btheta_i \mid \cdots)\right] \propto -\bY_i + m_i\bphi(\btheta_i) + \bSigma^{-1}(\btheta_i - \bmu - \bLambda\bff_i).
\end{equation}
For some given number of steps $n_{\text{steps}}$ and step size $\epsilon$, HMC proceeds as follows for each subject $\ind{i}{n}$ separately:
At each iteration, initialize a $K$-vector of ``position'' variables
$\bu = (u_1,\ldots,u_K)'$ at $\btheta_i$ and a $K$-vector of ``momentum'' variables
$\bv = (v_1,\ldots,v_K)'$ with independent random draws from a standard normal
distribution. Calculate the current ``total energy'' of the system
$H(\bu, \bv) = U(\bu) + K(\bv)$ where $U(\bu)$ is \eqref{eq:nlp_theta} evaluated
at $\bu$ and $K(\bv) = \frac{1}{2}\sum_{k=1}^K v_k^2$ mark the ``potential'' and
``kinetic'' energies, respectively. Furthermore, let $\nabla U(\bu)$ denote
\eqref{eq:grad_nlp_theta} evaluated at $\bu$.

Next we initiate $n_{\text{steps}}$ Leapfrog steps through our dynamic system.
\begin{enumerate}
\item Start with a half-step for momentum: $\bv \leftarrow \bv - \frac{\epsilon}{2}\nabla U(\bu)$.
\item Now, $n_{\text{steps}} - 1$ alternating full-steps for position and momentum: $\bu \leftarrow \bu + \epsilon \bv$, then $\bv \leftarrow \bv - \epsilon \nabla U(\bu)$.
\item End with one final full-step for position and a half-step for momentum: $\bu \leftarrow \bu + \epsilon \bv$, then $\bv \leftarrow \bv - \frac{\epsilon}{2}\nabla U(\bu)$.
\item Denote $\bu^* \leftarrow \bu$ and $\bv^*\leftarrow -\bv$ and accept the proposed $\bu^*$ as the new $\btheta_i$ draw with probability $\min\left\{1, \exp\left[H(\bu,\bv)-H(\bu^*,\bv^*)\right]\right\}$.
\end{enumerate}
The rate of acceptance depends on the values of $n_{\text{step}}$ and
$\epsilon$ so some degree of tuning is recommended during the burn-in phase of
MCMC to achieve a suitable acceptance rate (heuristically between 25\% and 45\%).

\subsection*{Gibbs sampling}\label{ssec:gibbs}

The remaining parameters in the model are amenable to Gibbs updates by sampling
from their full conditional distributions. The full conditionals are as follows:
\begin{equation*}
  \begin{aligned}[t]\label{eq:post_mu}
    \mu_k\mid\cdots &\sim N(\mu_{\mu_k}, \sigma^2_{\mu_k}) \for\ind{k}{K}\where \\
    \sigma^{-2}_{\mu_k} &= n\sigma^{-2}_k + \sigma^{-2}_{\mu} \and
    \mu_{\mu_k} = \sigma^{2}_{\mu_k} \left[\sigma^{-2}_k \sum\limits_{i=1}^n\left(\theta_{ik}-\sum_{l=1}^L\lambda_{kl}\ff_{il}\right)\right],
  \end{aligned}
\end{equation*}
\begin{equation*}
    \begin{aligned}[t]\label{eq:post_lambda}
      \blambda_{k}\mid\cdots &\sim N_{L}\left(\bmu_{\lambda_{k}}, \bSigma_{\lambda_{k}}\right) \for \ind{k}{K} \where \\
      \bSigma^{-1}_{\lambda_{k}} &= \sigma^{-2}_k\bF'\bF + \bPsi_{k}^{-1}  \and
      \bmu_{\lambda_{k}} = \bSigma_{\lambda_{k}}\bF'\bR_{k}\with \bF = (\bff_1,\ldots,\bff_n)',\\
      &\bPsi_{k} \text{ the $L\times L$ diagonal matrix with diagonal elements } \left(\psi_{kl}\xi_{kl}^2\tau_l^2\right)_{l=1}^L \and \\
      &\bR_{k} \text{ the $n$-vector with elements } \left[\sigma^{-2}_{k}(\theta_{ik} - \mu_{k})\right]_{i=1}^{n},
    \end{aligned}
\end{equation*}
\begin{equation*}
  \begin{aligned}[t]\label{eq:post_psi}
    \psi_{kl}^{-1}\mid\cdots &\sim \text{InvGau}\left(\mu_{\psi_{kl}}, 1\right) \for \ind{l}{L},\;\ind{k}{K} \where \mu_{\psi_{kl}} = \frac{\xi_{kl}\tau_l}{|\lambda_{kl}|}.
  \end{aligned}
\end{equation*}

Posterior updating of $\tau_l$ and $\bxi_l$ require sampling from the generalized inverse Gaussian
distribution, i.e., $U\sim \text{GIG}(\eta, \rho, \chi)$ with density
$f(u) \propto u^{\eta-1}\exp\left[-0.5(\rho u + \chi/u)\right]$, for which dedicated
algorithms exist \parencite{hormann2014generating}.
\begin{equation*}
  \begin{aligned}[t]\label{eq:post_tau}
      \tau_l\mid\cdots &\sim \text{GIG}\left(Ka_l - K, 2\nu, 2\sum_{k=1}^K\frac{|\lambda_{kl}|}{\xi_{kl}}\right) \for\ind{l}{L},
  \end{aligned}
\end{equation*}
\begin{equation*}
  \begin{aligned}[t]\label{eq:post_xi}
    \xi_{kl} &= \frac{T_{kl}}{T_{+l}} \where T_{kl}\mid\cdots \sim \text{GIG}\left(a_l-1,2\nu,2|\lambda_{kl}|\right) \and T_{+l}=\sum_{k=1}^KT_{kl} \\
    &\text{as derived in Theorem 2.2 of \textcite{bhattacharya2015dirichlet}},
  \end{aligned}
\end{equation*}
\begin{equation*}
  \begin{aligned}[t]\label{eq:post_nu}
      \nu\mid\cdots &\sim \text{Gam}\left(c_0 + K\sum_{l=1}^La_l, d_0 + \sum_{l=1}^L\tau_l\right),
  \end{aligned}
\end{equation*}
\begin{equation*}
  \begin{aligned}[t]\label{eq:post_f}
    \bff_i\mid\cdots &\sim N_L\left(\bmu_{\ff_i}, \bSigma_{\ff}\right) \for \ind{i}{n} \where \\
    \bSigma^{-1}_{\ff} &= \bLambda'\bSigma^{-1}\bLambda + \bI \and \bmu_{\ff_i} = \bSigma_{\ff}\left[\bLambda'\bSigma^{-1}(\btheta_i - \bmu) + (\bb\bx_i + \bg_{z_i})\right],
  \end{aligned}
\end{equation*}
\begin{equation*}
    \begin{aligned}[t]\label{eq:post_g}
        g_{rl}\mid\cdots &\sim N\left(\mu_{g_{rl}}, \sigma^{2}_{g_{rl}}\right) \for \ind{l}{L},\;\ind{r}{q} \where \\
      \sigma^{-2}_{g_{rl}} &= \sum_{i=1}^nI(z_i=r) + \sigma^{-2}_{g} \and \mu_{g_{rl}} = \sigma^{2}_{g_{rl}}\sum_{i=1}^n\left(\ff_{il} - \sum_{j=1}^p b_{jl}x_{ij}\right)I(z_i=r).
    \end{aligned}
\end{equation*}

For notational convenience,
let $\tilde{b}_{jl} = b_{jl}\omega_{jl}$ and
each column of $\tilde{\bX}_l = \bX * \diag\left(\omega_{1l}, \ldots, \omega_{pl}\right)$
is the corresponding column of $\bX$ if $\omega_{jl}=1$ and zero otherwise.
\begin{equation*}
    \begin{aligned}[t]\label{eq:post_b}
        \bb_l\mid\cdots &\sim N_p\left(\bmu_{b_l}, \bSigma_{b_l}\right) \for \ind{l}{L} \where \\
        \bSigma_{b_l}^{-1} &= \tilde{\bX}'_l\tilde{\bX}_l + \sigma_{b_l}^{-2}\bI \and \bmu_{b_l} = \bSigma_{b_l} \tilde{\bX}'_l\left(\bff_l-\bG_l\right) \with \bG_l = (g_{z_1 l}, \ldots, g_{z_n l})',
    \end{aligned}
\end{equation*}
\begin{equation*}
    \begin{aligned}[t]\label{eq:post_omega}
      Pr(\omega_{jl} = 1 \mid \cdots) &= \frac{\pi_{jl}^{(1)}}{\pi_{jl}^{(0)}+\pi_{jl}^{(1)}} \for \ind{l}{L},\;\ind{j}{p} \where \\
      \pi_{jl}^{(0)} &=(1-\pi_j)\exp\left[-\frac{1}{2}\sum_{i=1}^n\left(d_{il}^{\setminus j}\right)^2\right]  \and \pi_{jl}^{(1)} =\pi_j\exp\left[-\frac{1}{2}\sum_{i=1}^n\left(d_{il}^{\setminus j}-b_{jl}x_{ij}\right)^2\right],\\
      d_{il}^{\setminus j} &= \ff_{il}-g_{z_i l}-\sum_{\ell=1}^p\tilde{b}_{\ell l}x_{i\ell} + \tilde{b}_{jl}x_{ij},
    \end{aligned}
\end{equation*}
\begin{equation*}
  \begin{aligned}[t]\label{eq:post_pi}
  \pi_j\mid\cdots \sim \text{Beta}(a_0 + S_j, b_0 + L - S_j)\where S_j=\sum_{l=1}^L \omega_{jl},
  \end{aligned}
\end{equation*}
\begin{equation*}\label{eq:post_tau}
  \sigma_k^{2}\mid\cdots\sim \text{InvGam}\left[
    u_0 + \frac{n}{2},
    v_0 + \frac{1}{2}\sum\limits_{i=1}^n
    \left(\theta_{ik} - \mu_k - \sum_{l=1}^L\lambda_{kl}\ff_{il}\right)^2
    \right] \for \ind{k}{K},
\end{equation*}
\begin{equation*}\label{eq:post_mu_var}
  \sigma_\mu^{2}\mid\cdots\sim \text{InvGam}\left(u_0 + \frac{K}{2}, v_0 + \frac{1}{2}\sum_{k=1}^K \mu_k^2 \right),
\end{equation*}
\begin{equation*}\label{eq:post_delta_var}
  \sigma_g^{2}\mid\cdots\sim \text{InvGam}\left(
    u_0 + \frac{qL}{2},
    v_0 + \frac{1}{2}\sum_{l=1}^L\sum_{r=1}^q g_{rl}^2
    \right),
\end{equation*}
\begin{equation*}\label{eq:post_alpha_var}
  \sigma_b^{2}\mid\cdots\sim \text{InvGam}\left(
      u_0 + \frac{1}{2}\sum_{l=1}^L\sum_{j=1}^p\omega_{jl},
    v_0 + \frac{1}{2}\sum_{l=1}^L\sum_{j=1}^p\tilde{b}_{jl}^2
    \right).
\end{equation*}

\end{document}